\documentclass[useAMS,usenatbib,maxnames=4]{mn2e}
\usepackage{bibentry}
\usepackage{mathtools}
\usepackage{float}
\usepackage{color}
\usepackage{amssymb}
\usepackage{epsfig}
\newcommand{\HII}{\textsc{Hii}~}
\newcommand{\HI}{\textsc{Hi}~}
\newcommand{\CI}{\textsc{Ci}~}
\newcommand{\CII}{\textsc{Cii}~}

\newcommand{\gsim}{\lower.7ex\hbox{$\;\stackrel{\textstyle>}{\sim}\;$}}
\newcommand{\lsim}{\lower.7ex\hbox{$\;\stackrel{\textstyle<}{\sim}\;$}}

\newcommand{\apj}{ApJ}

\newcommand{\apjs}{ApJS}
\newcommand{\mnras}{MNRAS}
\newcommand{\aap}{A\&A}

\newcommand\textlcsc[1]{\textsc{\MakeLowercase{#1}}}
\title[\CI as a tracer of H$_2$ in the high-redshift Universe]{Atomic carbon as a powerful tracer of molecular gas in the high-redshift Universe: perspectives for ALMA}

\author[Tomassetti et al.]{M. Tomassetti$^{1}$\thanks{E-mail:mtomas@astro.uni-bonn.de\newline
Member of the International Max Planck Research School (IMPRS) for Astronomy and Astrophysics at the Universities of Bonn and Cologne}, C. Porciani$^{1}$, E. Romano-D\'{\i}az$^{1}$, A.~D. Ludlow$^{1}$, P.~P. Papadopoulos$^{2}$\\
  $^{1}$Argelander Institut f\"{u}r Astronomie, Auf Dem H\"{u}gel 71, Bonn D-53121, Germany\\
  $^{2}$School of Physics \& Astronomy, Cardiff University, The Parade, Cardiff CF24 3AA, UK}

\begin{document}

\date{Accepted 2014 August 19. Received 2014 August 19; in original form 2014 May 2}

\pagerange{\pageref{firstpage}--\pageref{lastpage}} \pubyear{2014}

\maketitle

\label{firstpage}

\begin{abstract}
  We use a high-resolution  hydrodynamic simulation that tracks the non-equilibrium
  abundance   of   molecular  hydrogen  within  a   massive
  high-redshift galaxy to produce mock Atacama Large Millimeter Array (ALMA) maps of the fine-structure
  lines  of atomic carbon,  \CI 1-0  and \CI  2-1.  Inspired  by recent
  observational and theoretical work, we assume that \CI is
  thoroughly mixed within giant molecular clouds and demonstrate that its emission is
  an excellent proxy for H$_2$. Nearly all of the
  H$_2$ associated with the galaxy can be  detected  
  at redshifts  $z<4$ using  a compact  interferometric
  configuration with  a large synthesized beam (that  does not resolve
  the   target  galaxy)   in   less  than 4 h  of   integration
  time. Low-resolution imaging  of the \CI lines (in  which the target
  galaxy is resolved into three to four beams) will detect $\sim$80 per cent of the H$_2$
  in  less than  12  h  of  aperture
  synthesis. In this case, the  resulting data cube also provides 
  the  crucial information necessary  for determining  the dynamical
  state of  the galaxy. We  conclude that ALMA observations  of the
  \CI  1-0 and  2-1 emission  are  well-suited for  extending the
  interval of  cosmic look-back time  over which the H$_2$ 
  distributions, the dynamical  masses, and the Tully-Fisher relation
  of galaxies can be robustly probed.
\end{abstract}


\section{Introduction}\label{sec:introduction}

Forbidden fine-structure lines of neutral atomic carbon can be used as tracers of the
molecular mass in galaxies \citep{Gerin+00,Papadopoulos+04}. The ground electronic state 
of neutral carbon, $\rm 1s^2\,2s^2\,2p^2$, is split into five fine-structure levels that, 
in spectroscopic notation, are denoted by $\rm ^3P_J$ (with $J=0,1$ and 2), $\rm ^1D_2$, and $\rm ^1S_0$. 
The excited fine levels $\rm ^3P_1$ and $\rm ^3P_2$ lie only 23.6 and 62.4 K above the ground 
state ($\rm ^3P_0$) and are therefore easily populated by particle collisions in the cold 
interstellar medium. Two magnetic-dipole transitions are allowed between the 
fine-structure levels: $\rm ^3P_1 \to {\rm ^3P_0}$ (which we refer to as \CI 1--0) has a rest 
frequency of 492.1607 GHz, while $\rm ^3P_2 \to {\rm ^3P_1}$ (\CI 2--1) produces electromagnetic 
radiation at 809.3435 GHz.

Early  one-dimensional  models   of  photon-dominated  regions  (PDRs)
confined the presence of \CI to a thin transition layer separating the
outer    ionized    zone     from    the    CO-rich    inner    volume
\citep{Kaufman+99}. However,  large-scale \CI  surveys of the  Orion A
and  B molecular clouds  \citep{Ikeda+02} and  of the  Galactic center
\citep{Ojha+01}  as well as observations of nearby galaxies \citep[][]{Israel+02,Zhang+14} have  found that the \CI  1--0, 2--1 line
emission is  fully concomitant  and strongly correlates  with $^{12}
\rm    CO$    intensity.    In   addition,    several    observations
\citep[e.g.][]{Frerking+89,Schilke+95,Kramer+08}  have  shown  that
\CI  is ubiquitous  in  giant molecular  clouds  (GMCs). Clumpy  PDR
models \citep[e.g.][]{Spaans+96} suggest  that the surface layers of
\CI  are  evenly  spread   across  GMCs. Moreover  turbulent
diffusion,  cosmic ray fluxes,  and  non-equilibrium chemistry  all
help to maintain a nearly constant $[\CI\!\!]/[\rm H_2]$ abundance
ratio throughout  most of  the mass of  a typical  molecular cloud
\citep{Papadopoulos+04}.  A recent  numerical  simulation of  a
turbulent  molecular  cloud confirms  that  \CI  emission should  be
widespread  in GMCs, with  most of  the neutral  carbon at gas
densities between $10^2$ and $10^4$ cm$^{-3}$ and kinetic temperatures of 
  $T_{\rm k}\sim$ 30 K \citep{Glover+14}.

Sensitivity  limitations   and   the   low  atmospheric
trasmissivity  in the  short sub-mm  regime made  the  detection and
imaging of \CI lines in the local Universe  difficult (hampering the realization
  that they  do not  conform to the  standard stratified PDR picture).
At redshifts $z\sim 2-4$, where  the \CI lines
are  redshifted  into  more  favourable atmospheric  windows,  only  a
handful of galaxies have been observed in the 1--0 and 2--1
transitions  \citep[][and references therein]{Weiss+03,
Weiss+05,Walter+11,Alaghband-Zadeh+13}. However, this sample will undoubtedly grow 
in size due to forthcoming observations from the Plateau de Bure Interferometer (possibly upgraded to the Northern
Extended Millimeter Array), the
Submillimeter Array, the Atacama Pathfinder
Experiment and the Atacama Large Millimeter Array (ALMA), as well as
with the advent of new facilities such as
and the Cerro Chajnantor Atacama Telescope. The
improved sensitivity of these instruments will yield observations that
significantly  enhance  our  understanding  of the  molecular  gas  content
of distant galaxies and of the history of cosmic star formation.

In  this  letter,  we   use  a fully cosmological simulation 
of the formation of a massive, high-redshift galaxy and track its non-equilibrium
H$_2$ abundance across cosmic time. Using this simulation, we produce mock ALMA
maps of the associated \CI 1-0  and \CI  2-1 fine-structure line emission, 
and demonstrate that such observations can be used to robustly estimate the 
molecular content of massive systems at high redshifts.

\section{Methods}\label{sec:methods}

The simulation follows the formation of a massive galaxy up to $z=2$ in its full 
cosmological context using the adaptive-mesh-refinement code \textlcsc{RAMSES} \citep{Teyssier+02}. It achieves
a physical resolution $\simeq \, 180$ pc, and includes gas cooling, star formation, 
feedback and metal enrichment from stellar evolution. We use a novel sub-grid 
treatment of the formation and destruction of molecular hydrogen within unresolved GMCs to 
track the non-equilibrium abundance of H$_2$. At $z=2$, the galaxy has a stellar mass of $7.7\times10^{10}$ M$_\odot$, a star-formation rate of 43 M$_\odot$ yr$^{-1}$
and an H$_2$ mass of $4.4\times10^{10}$ M$_\odot$, which are consistent with the properties of 
sub-mm galaxies for which \CI emission has been already detected
at $z>2$.
This simulation is described in full detail in \citet{Tomassetti+14},
to which we refer the reader for further details.

To produce \CI 1-0  and \CI 2-1 emission maps, we assume that atomic carbon is 
thoroughly mixed with molecular hydrogen and 
that the $[\CI\!\!]/[\rm H_2]$ abundance ratio is constant on kpc scales. (This quantity is 
observed to range between $10^{-5}$ and $10^{-4}$ in non-star-forming 
and star-forming clouds, respectively \citep{Frerking+89,Weiss+05,Walter+11,Danielson+11}). In the following, 
we adopt the representative value of $[\CI\!\!]/[\rm H_2]\simeq 3\times10^{-5}$ \citep{Alaghband-Zadeh+13}, but discuss in 
Section \ref{sec:results} how changes to this parameter affect our results.

At the frequencies of the \CI fine-structure lines our simulated galaxy is optically 
thin and the flux density per unit frequency observed on Earth, $S_{\CI\!\!}$, for the 
transition between levels u and l is
\begin{equation}\label{eq:flux_density}
S_{\CI\!\!} = \frac{A_{\rm ul}\,h\nu_{\rm ul}\,\Omega_{\rm B}\,Q_{\rm u}\overline{N}_{\CI\!\!,\Delta\nu}}{4{\rm \pi}\,(1+z)^4\Delta\nu}.
\end{equation}
Here  $A_{\rm ul}$  is the  Einstein  coefficient, $h$  is the Planck
constant,  $\nu_{\rm  ul}$   the  rest-frequency  of  the  transition,
$\Omega_{\rm B}$  the beam  size, $z$ the  redshift at which the  galaxy is observed and
$\Delta\nu$     the     frequency     bandwidth.    The     term
$\overline{N}_{\CI\!\!,\Delta\nu}$  denotes  the beam-averaged  column
density of carbon atoms with line-of-sight velocities corresponding to
the  frequency  range $\Delta\nu$  for  the  line transition.  Lastly,
$Q_{\rm  u}=n_{\rm u}/n_{\CI\!\!}<1$ is  the fractional  population of
level u,  where $n_{\CI\!\!}$  is the total  number density  of atomic
carbon. Given that  the critical densities of the  \CI transitions are
comparable  to the  densities of  GMCs,  it is  reasonable to  compute
$Q_{\rm u}$ assuming local  thermodynamic equilibrium\footnote{We have 
verified that excitations due to the cosmic microwave 
background alter the population ratios only at the per cent level at redshifts as high as 5.} (LTE).  
  We  assume  a  gas  kinetic
temperature  of $T_{\rm  k}=30$  K \citep
{Walter+11},  for  which  $Q_{\rm 1}\simeq0.46$  and  $Q_{\rm2}\simeq0.21$.  
The impact of varying $T_{\rm  k}$ is discussed in the next section.

Our goal  is to demonstrate the power of
using \CI\!\!-line ALMA observations to estimate the H$_2$ mass of high-$z$ galaxies.  Noise levels 
are computed using the ALMA sensitivity calculator\footnote{http://almascience.eso.org/proposing/sensitivity-calculator}
for an array configuration of fifty 12 m antennas. We
assume  that our  simulated  galaxy is  observed  at an elevation
of 45$^\circ$ and adopt a water vapour column density of $1.8$ mm.

\section{Results}\label{sec:results}

\begin{figure}\center
  \includegraphics[scale=0.35]{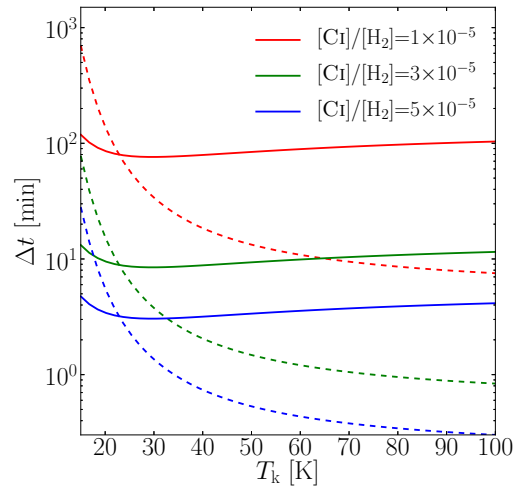}
  \caption{\label{fig:integration}Integration time, $\Delta t$, required to detect the
  \CI signal from our simulated galaxy at $z=2$ as
    a function of the kinetic gas temperature. We plot the time required to observe 
  the brightest
    pixel with a signal-to-noise ratio of five in the (continuum-subtracted) surface-brightness maps.  All curves refer to a compact ALMA configuration with a baseline length of $D=150$ m
    whose synthesized beam is larger than the simulated galaxy at the observed redshift. We assume that our  galaxy is observed at an elevation of 45$^\circ$. Solid and
    dashed lines refer to the 1-0 and 2-1 transitions, respectively.}
\end{figure}

\begin{figure*}
  \includegraphics[scale=0.25]{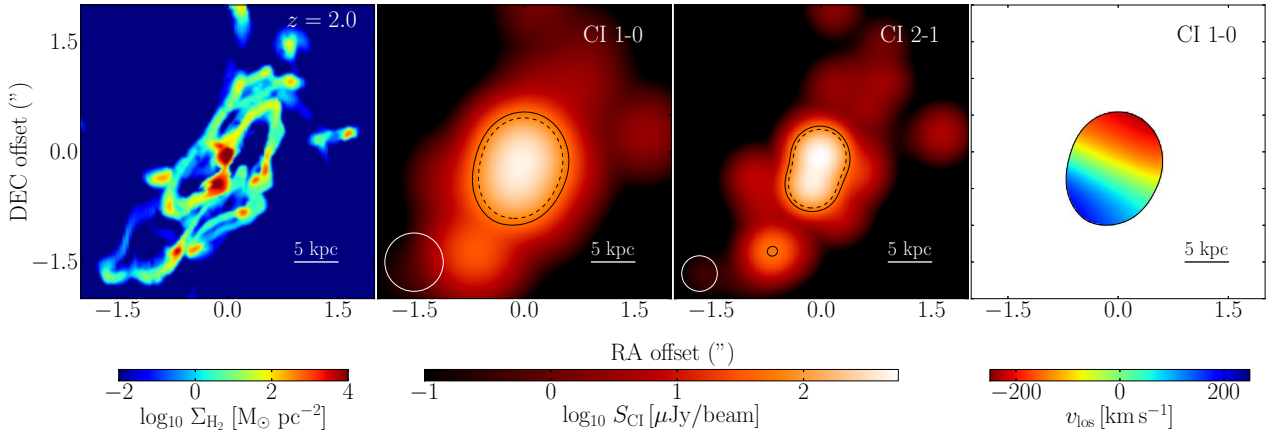}
  \caption{\label{fig:observations}Left: simulated H$_2$ surface density at
$z=2$. Middle:
flux density of the \CI 1--0 (middle-left) and 2--1 (middle-right)
transitions for the projection shown on
the left assuming a bandwidth $\Delta v=\pm2\sigma$ and a baseline
$D=700$ m. Contours are drawn at S/N levels of 3
(solid) and 5 (dashed) for an integration time of
$\Delta t=12$ h; white circles in the lower-left corners
indicate the beam shape and size.
Right: flux-weighted mean velocity map determined from the
\CI 1-0 signal using velocity channels of 50 km s$^{-1}$.
Contours show S/N=3 for $\Delta t=12$ h for the flux density in
the middle-left panel. Mock observations assume that our
galaxy is observed at an elevation of 45$^\circ$.}
\end{figure*}

For a detection experiment, we first consider a compact ALMA
configuration with a baseline length  of $D=150$ m.  In this case, the
synthesized beam is large (for observations at $z=2$, it has an FWHM of
$3.1$ and  $1.9$ arcsec for \CI  1--0 and 2--1,  respectively) and the
simulated galaxy  is not  spatially resolved at any  redshift $z>2$.   In Fig.
\ref{fig:integration}  we show,  as  a function  of  $T_{\rm k}$,  the
integration time required to detect the  galaxy at $z=2$.  We integrate the
data cube over the velocity range $\Delta v\simeq\pm 450$ km s$^{-1}$,
which corresponds to four  times the line-of-sight velocity dispersion, $\sigma$,
of the  H$_2$ gas within the beam. We compute  the integration
time by  requiring the brightest  pixel in the  (continuum subtracted)
surface-brightness map  to have a  signal-to-noise (S/N) ratio of 5. For
our fiducial  values of the kinetic temperature  and carbon abundance,
approximately 4 min of exposure is enough to detect the 2--1 line, 
whereas 8 min are needed  for   the  1--0  line.  These  values  scale  as
$([\CI\!\!]/[\rm  H_2])^{-2}\,Q_{\rm u}^{-2}$, suggesting  that 76 (34) min of  
integration is  necessary for an  abundance ratio  of $\sim 10^{-5}$ for the 1--0 (2--1) line.  Note that for high values of
the gas temperature, the 2--1 line is brighter than the 1--0 line. However, it may be difficult  
to  detect for $T_{\rm  k}\ll  30$ K  as  the  $\rm ^3P_2$ level  may not be significantly populated.

Using equation (\ref{eq:flux_density}), we can estimate the H$_2$ mass contained 
  within $\Omega_{\rm B}$, from the flux in the brightest pixel:
\begin{equation}\label{eq:mass_max_pixel}
M_{\rm H_2}^{\rm max}=\frac{4{\rm \pi}\,m_{\CI\!\!}\Delta v\,S_{\CI\!\!}^{\rm max}\,D_{\rm L}^2}{6\,[\CI\!\!]/[{\rm H_2}]\,A_{\rm ul}\,Q_{\rm u}\,h\, c\,(1+z)}.
\end{equation}
Here $S_{\CI\!\!}^{\rm max}$ is the pixel brightness, $m_{\CI\!\!}$ is the mass of a carbon atom, $D_{\rm L}$ 
the luminosity distance and $c$ is the speed of light. For our assumed values of the \CI temperature and the $[\CI\!\!]/[\rm H_2]$ 
abundance ratio, this gives $4.3\times10^{10}$ M$_\odot$ and $4.0\times10^{10}$ M$_\odot$ for the 1--0 and 2--1 
transitions, respectively.

In  practice the  kinetic  temperature can  be  determined using  the
line-intensity   ratio   \CI    2--1/\CI   1--0  under the assumption of  LTE
\citep[e.g.][]{Stutzki+97}. One can  use also the  non-LTE
$Q_{\rm u}(n,T_{\rm k}$) expressions  (Papadopoulos et al. 2004), given certain
constraints on the average gas density exist (e.g. via multi-$J$ CO and
HCN line observations). However, the relative \CI
abundance must be calibrated  against multiple observations of 
other  emission lines  as well as  measures  of the  dynamical  mass in  GMCs
(e.g. from CO). Observational  noise together with uncertainties in $T_{\rm k}$
and $[\CI\!\!]/[\rm H_2]$ will, of course, generate scatter around our
ideal results.

The spectrum extracted from the data cube at the location of the brightest pixel can be used to obtain
 $\sigma$ and the dynamical mass of the galaxy. Over the redshift range $2<z<3$, 
we find that dynamical masses, estimated as $M_{\rm dyn}=\alpha\,\sigma^2\,R_{\rm B}/G$ (where $G$ is 
the gravitational constant, $R_{\rm B}=\rm FWHM/(2\,\sqrt{\ln 2})$ the beam radius, and $\alpha=3.4$; \citealt{Erb+06}),
are within a factor of two of the {\textit{total}} mass within a spherical aperture of radius $R_{\rm B}$ centred on 
the galaxy.

Next, we construct low-resolution ALMA maps of the H$_2$ distribution within our simulated galaxy assuming a baseline 
length of $D=700$ m. At $z=2$, the galaxy is then resolved by three to four telescope beams (the FWHM of the ALMA beam is 
$0.7\,(5.6)$ and $0.4\,(3.4)$ arcsec (kpc) for the 1--0 and 2--1 transitions, 
respectively) and the noise level for an integration time $\Delta t$ and a velocity bandwidth of roughly $\pm350$
km s$^{-1}$ is $\sigma_{\rm rms}\simeq 50\,(\Delta t/\rm 1\,h)^{-1/2}$ $\rm \mu$Jy beam$^{-1}$. The results are shown in Fig. \ref{fig:observations}.
In the left-hand panel, we plot the H$_2$ surface density at the spatial resolution of the 
simulation (for a line of sight that forms an angle of 45$^\circ$ with respect to the disk). 
The two middle panels show the corresponding \CI flux densities for the 1--0 (middle-left) and 2--1 (middle-right) transitions. 
Detecting the brightest pixel in the maps with an S/N ratio
of 3 (5) requires an integration time of $\sim 7$ (20) min
for the 1--0 transition and $\sim 5$ (14) min for 2--1.
S/N contours drawn for an integration time of 12 h reveal 
that the detectable signal originates from regions where the beam-averaged H$_2$ column density 
is $\gsim \, 70$ M$_\odot$ pc$^{-2}$.
In the right-hand panel, we use the data cube to draw a map of the mean line-of-sight velocity. Our results show that
integration times on the order of 1 h are sufficient to 
retrieve detailed dynamical information.

\begin{figure}\center
  \includegraphics[scale=0.35]{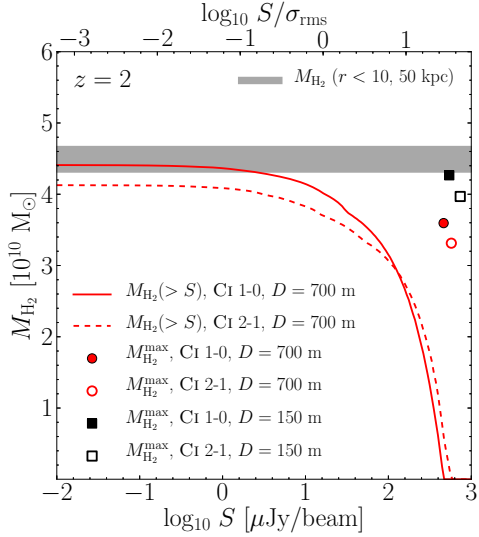}
  \caption{\label{fig:recovered} Red lines show the H$_2$ mass obtained by integrating over 
    the solid angle subtended by regions over which $S_{\CI\!\!}$ 
    is above a given flux limit, $S$, for the 1--0 (solid line) and 2--1 (dashed line) transitions. 
    The grey shaded region shows the H$_2$ mass that lies within $10$ and $50$ kpc 
    from the centre of the galaxy at $z=2$.
    Symbols indicate the flux in the brightest pixel and the corresponding H$_2$ mass contained within 
    $\Omega_{\rm B}$ as inferred from equation (\ref{eq:mass_max_pixel}) for the 1--0 (filled symbols) and 2--1 
    (open symbols) transitions. Results are plotted for the $D=700$ m (circles) 
    and $D=150$ m (squares) cases. The upper $x$-axis shows the corresponding S/N ratio for 
    the 1--0 transition (and integration time $\Delta t=12$ h). For this plot we have adopted
    $[\CI\!\!]/[\rm H_2]=3\times10^{-5}$ and $T_{\rm k}=30$ K, and assumed an elevation of 45$^\circ$.}
\end{figure}

It is useful to estimate how accurately we can reconstruct the H$_2$ mass of the galaxy from resolved 
imaging of the \CI 1--0 and 2--1 lines. To do so, we compute the H$_2$ mass that 
occupies the same volume as the observable carbon atoms. This is achieved by convolving the H$_2$ 
surface-density maps with the ALMA beam and integrating over the solid angle subtended by regions 
in which $S_{\CI\!\!}$ is above a given flux limit, $S$. In  Fig. \ref{fig:recovered}, 
we show our results for the galaxy at $z=2$. 
The measured H$_2$ mass, $M_{\rm H_2}(>S)$, increases with decreasing the flux limit. For an integration time of 
12 h, the H$_2$ mass recovered from the area over which $S_{\CI\!\!}$ is above 3$\,\sigma_{\rm rms}$ 
is $3.6\times10^{10}$ M$_\odot$ for the 1--0 line and $3.3\times10^{10}$ M$_\odot$ for the 2--1 transition. 
Longer integrations, or reductions in the baseline length, boost the
signal from low-density regions. For $D=400$ m, for example, one finds $4\times10^{10}$ 
M$_\odot$ (for \CI 1--0) and $3.9\times10^{10}$ M$_\odot$ (for \CI 2--1).  These values can be directly 
compared to the H$_2$ mass of the galaxy: the grey shaded area in Fig. \ref{fig:recovered} shows the 
H$_2$ mass contained within 10 and 50 kpc from its center.

\begin{figure}\center
  \includegraphics[scale=0.35]{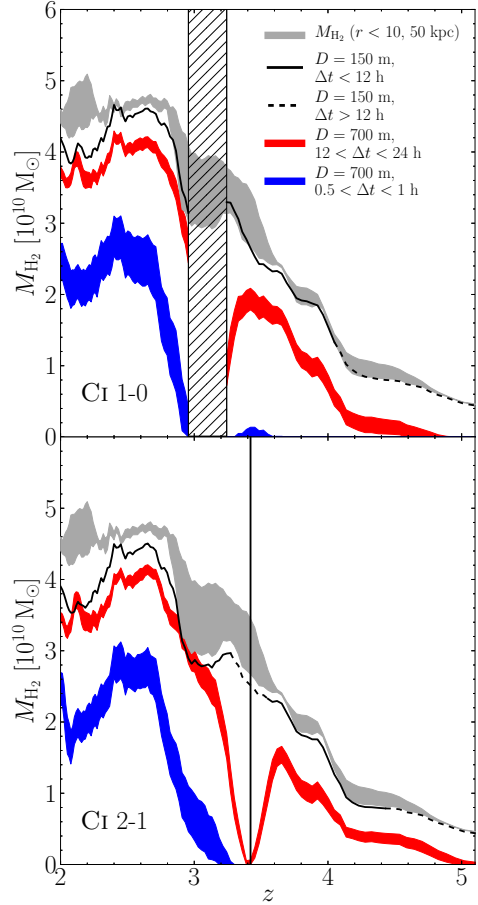}
  \caption{\label{fig:high_z}Molecular-hydrogen mass associated with our simulated galaxy as a function
    of redshift. The grey shaded region indicates the true underlying H$_2$ mass that is contained within
    spheres of radius 10 and 50 kpc centred upon the galaxy. The black line shows $M_{\rm H_2}^{\rm max}$ inferred 
    from the brightest pixel in our mock ALMA maps for a baseline length of $D=150$ m; dashed segments
    indicate redshifts                                   
    at which the integration time required to detect the peak flux above $5\,\sigma_{\rm rms}$ exceeds 12 h. Blue and red regions
    correspond to the H$_2$ mass, $M_{\rm H_2}(>S)$, contained in regions where the \CI signal is above $3\,\sigma_{\rm rms}$ as a function of redshift for $D=700$ m and integration times ranging from 12 to 24 h 
    (red) and from 30 min to 1 h (blue). 
    The hatched region in the top panel indicates the gap between bands 3 and 4 in the ALMA 
    setup, while the vertical line in the bottom panel marks the strong atmospheric water line at 183 GHz.
    At any redshift, we assume our galaxy to be observed at an elevation of 45$^\circ$.}
\end{figure}

The redshift dependence of the estimated H$_2$ mass can be obtained by 
following the galaxy backward in time and imaging a region surrounding it. 
The grey shaded areas in Fig. \ref{fig:high_z} show the evolution 
of the H$_2$ mass contained within 10 and 50 kpc from the galaxy centre (thicker shades indicate the presence of satellite galaxies),
which can be compared to the H$_2$ mass measured from the ALMA data. 
The black line in Fig. \ref{fig:high_z} corresponds to the H$_2$ mass, $M_{\rm H_2}^{\rm max}$, recovered from \CI observations  for the compact baseline of 150 m. For the 1--0 (2--1) line, the majority of the molecular 
hydrogen within the galaxy can be detected at redshifts as high as $z\sim4$ (3.3) with an integration 
time of less than 4 h. At $z\gsim 4$, the detection of \CI becomes more difficult, resulting in a lower fraction of the underlying H$_2$ mass being measured. This is due to the intrinsically 
lower H$_2$ mass of the galaxy at higher $z$, but also due to the surface-brightness dimming of sources at cosmological distances.
The blue shaded region in Fig. \ref{fig:high_z}
shows the estimated H$_2$ masses, $M_{\rm H_2}(>3\,\sigma_{\rm rms})$, obtained for 30 to 60 min of aperture synthesis
for a baseline length of 700 m. 
For $2<z<2.7$, these observations recover from 40 to 70 per cent of the
true H$_2$ mass in the innermost 10 kpc.
Longer integration times are required to probe the bulk of the 
molecular mass. The red region, for example, shows $M_{\rm H_2}(>3\,\sigma_{\rm rms})$ for $12<\Delta t<24$ h of 
integration with $D=700$ m. With this setup, at $z<3$, it is possible to recover nearly 90 per cent
of the H$_2$ mass within 10 kpc from the centre of the galaxy, while
about half of the underlying molecular mass can be identified up to
$z\lsim 4$.

\section{Conclusions}\label{sec:discussion}
We have used a high-resolution hydrodynamic simulation that follows the formation and evolution 
of a massive  galaxy at $z>2$ to construct template ALMA observations of the 
fine-structure \CI emission lines. Our simulation includes a sophisticated algorithm to track the 
non-equilibrium abundance of molecular hydrogen, accounting for its creation on dust grains and 
destruction by the Lyman-Werner photons emitted by young stars.

Assuming that \CI is well-mixed with H$_2$, as suggested
by recent theoretical models and observations, an abundance ratio of $[\CI\!\!]/[\rm H_2]=3\times10^{-5}$
and a gas temperature $T_{\rm k}$ = 30 K, we find that ALMA, in its
most compact configuration ($D = 150$ m), should be able to
detect our $z=2$ galaxy in only a few minutes of integration
time. In addition, an integration time of 1 hr will be sufficient to detect the \CI emission coming from galaxies similar
to ours at redshifts $z<3$. At higher $z$, longer integrations
are required to detect the bulk of the molecular content of
our galaxy. For example, for the 1--0 line, an integration
time of $\Delta t = 4$ h can recover 96 per cent of the molecular
mass that, at $z = 4$, lies within 10 kpc of the centre of the
galaxy. Higher-resolution observations (with $D = 700$ m and
$\Delta t = 12$ h) can be used to map a significant fraction (80-90 per cent) of
the underlying H$_2$ mass at redshifts as high as $z \sim 3$ and the densest
molecular regions (which contain nearly 50 per cent of the H$_2$ mass)
up to $z\sim 4.5$. The fine-structure \CI 
lines can therefore be used as a reliable tracer of molecular hydrogen in massive galaxies at high-redshift.

Using  the   \CI  fine-structure  lines   as  tracers  of   the  H$_2$
distribution has several advantages over traditional studies
employing rotational  transitions of  carbon monoxide. First,  the \CI
lines have a strong positive  \textit{K}-correction with respect to the low-$J$
CO transitions but similar excitation characteristics allowing
them to probe the bulk  of the H$_2$ mass \citep{Papadopoulos+04}.  On
the other hand, the two  high-$J$ CO lines with similar frequencies to
the \CI lines (CO $J$=4--3,  7--6), while often bright in star-forming
galaxies, trace {\textit{ only}} the dense and warm H$_2$ gas in star-forming
regions. Observations of these emission lines therefore enclose much more
compact regions of a galaxy and do not reliably sample the entire  H$_2$  
gas distribution.

Recent ALMA observations have detected \CII emission
in a gaseous starbursting disk at $z\sim 5$ \citep{DeBreuck+14}.  Its emission, while much
brighter than CI, is not tied  to the H$_2$ gas alone but also to
the  \HI and  \HII  distributions, which  are  currently not well constrained
in high-redshift galaxies (and will remain so until the completion of the Square Kilometre Array).
In addition, at $2.1<z<2.8$, when the cosmic star-formation rate peaks, \CII cannot be 
detected from the ground and observations at lower redshift are challenging even with ALMA.
We  note that high-fidelity, phase-stable imaging of
gas-rich  disks  at  high redshift is  necessary  for  tasks  such  as
determining dynamical masses.
The  two lower-frequency \CI  lines therefore carry several  advantages when used as
to estimate the  H$_2$ content and the dynamical-masses of galaxies  across cosmic
time.   Furthermore, when  combined with transitions  that trace
only dense  star-forming gas  (e.g.  CO 4--3,  HCN 1--0),  they 
also provide a powerful  probe of the star-formation mode (isolated
disk versus merger-driven) in the Universe \citep{Papadopoulos+12}.
We conclude that using  the two \CI lines (rather than low-$J$ CO
lines) opens up  a much larger fraction of  cosmic look-back time over
which the H$_2$  gas mass distribution, the dynamical  masses, and the
Tully-Fisher relation of galaxies can be accurately measured.

\section*{Acknowledgements}
We  thank Nadya  Ben Bekhti  for fruitful  discussions and an anonymous 
referee for a useful report. 
This work has been carried out within the Collaborative Research Centre 956, 
sub-project C4, funded by the Deutsche Forschungsgemeinschaft (DFG). MT was 
supported through  a stipend from the IMPRS in  Bonn and PPP was supported 
through an Ernest Rutherford Fellowship  from  STFC.   We  acknowledge  that 
the  results  of  this research  have   been  achieved  using  the   PRACE-2IP  
project  (FP7 RI-283493)  resources  HeCTOR based  in  the  UK  at the  UK  
National Supercomputing Service and the  Abel Computing Cluster based in Norway
at the University of Oslo.

\bibliographystyle{mn2e}


\begin{thebibliography}{}

\bibitem[\protect\citeauthoryear{{Alaghband-Zadeh}}{{Alaghband-Zadeh et al.}}{2013}]{Alaghband-Zadeh+13}
{Alaghband-Zadeh} S. et~al.,  2013, \mnras, 435, 1493

\bibitem[\protect\citeauthoryear{{Danielson}}{{Danielson et al.}}{2011}]{Danielson+11}
{Danielson} A.~L.~R. et~al.,  2011, \mnras, 410, 1687

\bibitem[\protect\citeauthoryear{{De Breuck}}{{De Breuck et al.}}{2014}]{DeBreuck+14}
{De Breuck} C. et~al.,  2014, \aap, 565, A59

\bibitem[\protect\citeauthoryear{{Erb}, {Steidel}, {Shapley}, {Pettini},
  {Reddy} \& {Adelberger}}{{Erb} et~al.}{2006}]{Erb+06}
{Erb} D.~K.,  {Steidel} C.~C.,  {Shapley} A.~E.,  {Pettini} M.,  {Reddy} N.~A.,
     {Adelberger} K.~L.,  2006, \apj, 646, 107

\bibitem[\protect\citeauthoryear{{Frerking}, {Keene}, {Blake} \&
  {Phillips}}{{Frerking} et~al.}{1989}]{Frerking+89}
{Frerking} M.~A.,  {Keene} J.,  {Blake} G.~A.,    {Phillips} T.~G.,  1989,
  \apj, 344, 311

\bibitem[\protect\citeauthoryear{{Gerin} \& {Phillips}}{{Gerin} \&
  {Phillips}}{2000}]{Gerin+00}
{Gerin} M.,  {Phillips} T.~G.,  2000, \apj, 537, 644

\bibitem[\protect\citeauthoryear{{Glover}, {Clark}, {Micic} \&
  {Molina}}{{Glover} et~al.}{2014}]{Glover+14}
{Glover} S.~C.~O.,  {Clark} P.~C.,  {Micic} M.,    {Molina} F.,  2014, preprint (arXiv:1403.3530)

\bibitem[\protect\citeauthoryear{{Ikeda}, {Oka}, {Tatematsu}, {Sekimoto} \&
  {Yamamoto}}{{Ikeda} et~al.}{2002}]{Ikeda+02}
{Ikeda} M.,  {Oka} T.,  {Tatematsu} K.,  {Sekimoto} Y.,    {Yamamoto} S.,
  2002, \apjs, 139, 467

\bibitem[\protect\citeauthoryear{{Israel} \& {Baas}}{{Israel} \&
  {Baas}}{2002}]{Israel+02}
{Israel} F.~P.,  {Baas} F.,  2002, \aap, 383, 82

\bibitem[\protect\citeauthoryear{{Kaufman}, {Wolfire}, {Hollenbach} \&
  {Luhman}}{{Kaufman} et~al.}{1999}]{Kaufman+99}
{Kaufman} M.~J.,  {Wolfire} M.~G.,  {Hollenbach} D.~J.,    {Luhman} M.~L.,
  1999, \apj, 527, 795

\bibitem[\protect\citeauthoryear{{Kramer}}{{Kramer et al.}}{2008}]{Kramer+08}
{Kramer} C. et~al.,  2008, \aap, 477, 547

\bibitem[\protect\citeauthoryear{{Ojha}}{{Ojha et al.}}{2001}]{Ojha+01}
{Ojha} R. et~al.,  2001, \apj, 548, 253

\bibitem[\protect\citeauthoryear{{Papadopoulos} \& {Geach}}{{Papadopoulos} \&
  {Geach}}{2012}]{Papadopoulos+12}
{Papadopoulos} P.~P.,  {Geach} J.~E.,  2012, \apj, 757, 157

\bibitem[\protect\citeauthoryear{{Papadopoulos}, {Thi} \&
  {Viti}}{{Papadopoulos} et~al.}{2004}]{Papadopoulos+04}
{Papadopoulos} P.~P.,  {Thi} W.-F.,    {Viti} S.,  2004, \mnras, 351, 147

\bibitem[\protect\citeauthoryear{{Schilke}, {Keene}, {Le Bourlot}, {Pineau des
  Forets} \& {Roueff}}{{Schilke} et~al.}{1995}]{Schilke+95}
{Schilke} P.,  {Keene} J.,  {Le Bourlot} J.,  {Pineau des Forets} G.,
  {Roueff} E.,  1995, \aap, 294, L17

\bibitem[\protect\citeauthoryear{{Spaans}}{{Spaans}}{1996}]{Spaans+96}
{Spaans} M.,  1996, \aap, 307, 271

\bibitem[\protect\citeauthoryear{{Stutzki}}{{Stutzki et al.}}{1997}]{Stutzki+97}
{Stutzki} J. et~al.,  1997, \apj, 477, L33

\bibitem[\protect\citeauthoryear{{Teyssier}}{{Teyssier}}{2002}]{Teyssier+02}
{Teyssier} R.,  2002, \aap, 385, 337

\bibitem[\protect\citeauthoryear{{Tomassetti}, {Porciani}, {Romano-Diaz} \&
  {Ludlow}}{{Tomassetti} et~al.}{2014}]{Tomassetti+14}
{Tomassetti} M.,  {Porciani} C.,  {Romano-Diaz} E.,    {Ludlow} A.~D.,  2014,
  preprint (arXiv:1403.7132)

\bibitem[\protect\citeauthoryear{{Walter}, {Wei{\ss}}, {Downes}, {Decarli} \&
  {Henkel}}{{Walter} et~al.}{2011}]{Walter+11}
{Walter} F.,  {Wei{\ss}} A.,  {Downes} D.,  {Decarli} R.,    {Henkel} C.,
  2011, \apj, 730, 18

\bibitem[\protect\citeauthoryear{{Wei{\ss}}, {Downes}, {Henkel} \&
  {Walter}}{{Wei{\ss}} et~al.}{2005}]{Weiss+05}
{Wei{\ss}} A.,  {Downes} D.,  {Henkel} C.,    {Walter} F.,  2005, \aap, 429,
  L25

\bibitem[\protect\citeauthoryear{{Wei{\ss}}, {Henkel}, {Downes} \&
  {Walter}}{{Wei{\ss}} et~al.}{2003}]{Weiss+03}
{Wei{\ss}} A.,  {Henkel} C.,  {Downes} D.,    {Walter} F.,  2003, \aap, 409,
  L41

\bibitem[\protect\citeauthoryear{{Zhang}}{{Zhang et al.}}{2014}]{Zhang+14} Zhang, Z.-Y. et~al.\ 2014, preprint (arXiv:1407.1444)


\end{thebibliography}

\label{lastpage}

\end{document}